\documentclass[prl,twocolumn,showpacs]{revtex4}
\usepackage{graphicx,amssymb,amsmath}

\newcommand{\Eq}[1]{Eq.~\eqref{#1}}
\newcommand{\eq}[1]{\eqref{#1}}

\newcommand{\beq}{\begin{equation}}
\newcommand{\eeq}{\end{equation}}

\newcommand{\beqa}{\begin{eqnarray}}
\newcommand{\eeqa}{\end{eqnarray}}

\newcommand{\Beqa}{\begin{eqnarray*}}
\newcommand{\Eeqa}{\end{eqnarray*}}
\newcommand{\nn}{\nonumber}

\newcommand{\pdag}{{\phantom{\dagger}}}

\DeclareMathOperator{\sign}{sgn}
\DeclareMathOperator{\im}{Im}
\DeclareMathOperator{\tsum}{\textstyle\sum}

\newcommand{\ket}[1]{\left\lvert{#1}\right\rangle}

\newcommand{\PRL}[3]{Phys. Rev. Lett.~\textbf{#1}, #2 (#3)}
\newcommand{\PRB}[3]{Phys. Rev. B~\textbf{#1}, #2 (#3)}

\newcommand{\JPCM}[3]{J. Phys. Condens. Matter~\textbf{#1}, #2 (#3)}
\newcommand{\NPB}[3]{Nucl. Phys. B~\textbf{#1}, #2 (#3)}
\newcommand{\JMP}[3]{J. Math. Phys.~\textbf{#1}, #2 (#3)}

\newcommand{\etal}{\textit{et al.}} 

\begin{document}

\title{Dynamic structure factor of the Calogero-Sutherland model}

\author{Michael Pustilnik}
\affiliation{School of Physics, Georgia Institute of Technology, 
Atlanta, GA 30332}


\begin{abstract}
We evaluate the dynamic structure factor $S(q,\omega)$ of a 
one-dimensional quantum Hamiltonian with the inverse-square 
interaction (Calogero-Sutherland model). For a fixed small $q$, 
the structure factor differs from zero in a finite interval of frequencies 
of the width $\delta\omega\propto q^2\!/m$. At the borders of this 
interval $S(q,\omega)$ exhibits power-law singularities with exponents 
depending on the interaction strength. The singularities are similar 
in origin to the well-known Fermi-edge singularity in the x-ray absorption 
spectra of metals. 
\end{abstract}

\maketitle

Fermi liquid theory proved to be extremely successful in describing 
interacting fermions~\cite{Nozieres}. The low-energy excitations 
of a normal Fermi liquid (FL) are classified the same way as 
the excitations of a reference system - a non-interacting Fermi 
gas. A weak residual interaction leads to a finite decay rate $1\!/\tau$
of FL's quasiparticles. Formally, the rate can be defined as the width
of the quasiparticle peak in the spectral function (imaginary part of a 
single-particle retarded Green function) $A(q,\omega)$. For FL, 
$A(q,\omega)$ at $T=0$ is a Lorentzian. Its width is easily evaluated 
in the second order of perturbation theory, which yields 
$1\!/\tau\propto\omega^2\!/\epsilon_F$, where $\omega(q)$ is the 
quasiparticle energy. 

It is well-known, however, that in one dimension (1D) even a weak 
interaction breaks down the FL description (see~\cite{1D_books} 
for recent reviews). A guide to understanding the properties of  
interacting 1D systems is provided by the Tomonaga-Luttinger 
model (TLM)~\cite{TL}, which plays the same role for the concept 
of the Luttinger liquid~\cite{Haldane} as the Fermi gas does for FL.
The TLM assumes a strictly linear fermionic dispersion relation. 
With this assumption, the TLM Hamiltonian can be diagonalized 
exactly~\cite{TL}, no matter how strong the interactions are. The 
corresponding elementary excitations are bosons, quanta of the 
waves of fermionic density. These bosons do not interact, have an 
infinite lifetime, and propagate without dispersion, $\omega(q) = uq$ 
(here $u$ is the plasma velocity). 
Therefore, a measurable quantity, the dynamic structure factor 
(density-density correlation function)~\cite{structure_factor}
\beq
S(q,\omega) = \int\!dx\,dt\,e^{i(\omega t - qx)} 
\bigl\langle\rho(x,t)\rho(0,0)\bigr\rangle,
\label{1}
\eeq
has an infinitely sharp peak,
$S_\text{TLM}(q,\omega)\propto q\delta (\omega - uq)$.

Applications of TLM to the description of ``real'' 1D fermions rely 
on the expansion of single-particle energies about the Fermi 
points $\pm \,p_F$~\cite{1D_books,Haldane},
\beq
\xi_k = \pm\,v_F k + k^2\!/2m +\ldots,
\quad
k=p\mp p_F,
\label{2}
\eeq
where the upper/lower sign corresponds to the right/left movers
(throughout this Letter we use units with $\hbar =1$). The linear in $k$ 
term in \Eq{2} is accounted for in TLM. The $k^2$ term 
generates interaction between the bosons with the coupling constant 
$\propto 1\!/m$~\cite{Haldane}, which broadens the peak in 
$S(q,\omega)$. However, in contrast with FL theory, this 
broadening is inaccessible by perturbation theory.

This can be seen by considering a special limit of LL, that of 
non-interacting fermions with spectrum \eqref{2}. In this case
the structure factor differs from zero only if $\omega$ lies within 
a finite interval 
$ 
\omega_-<\omega<\omega_+, 
$
where $\omega_\pm(q) = v_F q \pm\,q^2\!/2m$ for $q<2p_F$. 
Within this interval $S(q,\omega)$ is constant, $S = m/q$. Thus, 
at $q\to 0$ the structure factor indeed approaches the TLM form 
(with $u=v_F$), but in a very peculiar fashion: the peak in $S(q,\omega)$ 
at a fixed $q$ has a manifestly non-Lorentzian ``rectangular'' shape with the 
width $\delta\omega = q^2\!/m$.
Even this simple result is non-perturbative in the bosonic representation: 
the first-order in $1/m$ contribution to the boson's self-energy vanishes, 
while the next one diverges on the mass shell~\cite{Samokhin}.

Recently it was argued ~\cite{1DEG} that in the presence of interactions
between the fermions the shape of the peak in $S(q,\omega)$ remains 
to be non-Lorentzian. Moreover, even a weak interaction transforms 
the discontinuities at $\omega=\omega_\pm$ into power-law 
singularities. Here we approach the problem of finding $S(q,\omega)$ 
from the perspective of the exactly solvable Calogero-Sutherland 
model (CSM). 

The CSM Hamiltonian reads~\cite{Sutherland} 
\beq
H = -\sum_{i=1}^N\frac{1}{2m}\frac{\partial^2}{\partial x_i^2} 
+ \sum_{i<j} V(x_i-x_j),
\label{4} 
\eeq
where $V(x)$ is a periodic version of $1/x^2$ potential,  
\beq
V(x) = \frac{\lambda(\lambda - 1)\!/m}
{(L/\pi)^2\sin^2\!\bigl(\pi x/L\bigr)}\,.
\label{5} 
\eeq

The excitations of CSM can be described in terms of 
\textit{quasiparticles} and \textit{quasiholes}~\cite{Sutherland}. 
Quasiparticles are characterized by velocities $v$ in the range $|v|>u$, 
and an inertial mass $m$, the bare mass that enters \Eq{4}. Quasiholes 
have velocities $\bar v$ in the range $|\bar v|<u$, and fractional inertial 
mass $\bar{m} = m/\lambda$. The plasma velocity $u$ 
is given by~\cite{Sutherland}
\beq
u = \pi\lambda\rho_0/m,
\quad
\rho_0 = N/L.
\label{6}
\eeq

The momentum and energy (relative to the ground state) of an excited 
state of CSM characterized by a certain set of 
velocities 
$\{v_i,\bar v_j\}$ read \cite{Sutherland}
\beqa
P &=& \sum_i m v_i - \sum_j \bar{m} \bar v_j  
\label{7}\\
E &=& \sum_i \frac{1}{2}\, m\bigl(v_i^2-u^2\bigr) 
+ \sum_j\frac{1}{2}\,\bar{m} \bigl(u^2-\bar v_j^2\bigr).
\label{8}
\eeqa

\begin{figure}[h]
\includegraphics[width=0.8\columnwidth]{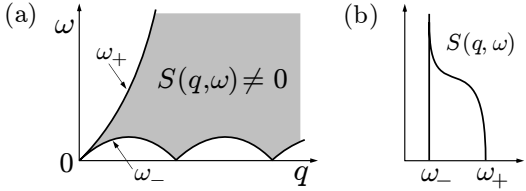}
\caption{
(a) The structure factor $S(q,\omega)$ differs from zero in a 
finite interval of frequencies 
$\omega_{-}\!\!<\!\omega\!<\!\omega_{+}$. At the borders of this interval
$S(q,\omega)$ exhibits power-law singularities,
$S\propto |\omega-\omega_\pm|^{\lambda^{\pm 1}-1}$, 
see Eqs. \eq{22} and \eq{23}. 
The low-energy $(\omega_-\to 0)$ sectors correspond 
to $q=2\pi\rho_0 I$, where $I$ is an integer.
(b) Dependence of $S(q,\omega)$ on $\omega$ at a fixed $q\neq 2\pi\rho_0 I$
and for a repulsive interaction $(\lambda>1)$. 
\label{Fig1}
}
\end{figure}

The result of the action of the local density operator 
\[
\rho^\dagger_q = \int_0^L\!dx\,e^{iqx}\rho(x),
\quad
\rho(x) = \frac{1}{L}\sum_{i=1}^N \delta(x-x_i)-\rho_0
\]
on CSM's ground state $\ket{0}$ has a remarkable property~\cite{Ha}: 
$\rho^\dagger_q \ket{0}$ is an eigenstate of CSM in which the velocities 
of all quasiparticles point in the same direction (positive for $q>0$).
This has a profound effect on the structure factor: $S(q,\omega)\neq 0$ 
only in a finite interval of frequencies, 
$\omega_-(q) <\omega<\omega_+(q)$, see Fig.~\ref{Fig1}(a) 
(the upper bound $\omega_+$ would be absent in a generic 1D 
system~\cite{1DEG}).

From this point on, we consider the rational values of $\lambda$ 
only, $\lambda = r/s$, where $r$ and $s$ are co-primes. 
In this case, the state  $\ket{R}=\rho^\dagger_{q>0} \ket{0}$ 
has exactly $s$ right-moving quasiparticles and exactly $r$ 
quasiholes \cite{Haldane_conjecture,ZH,Ha}. This is the simplest 
possible excitation that conserves the total inertial mass, 
\beq
s m - r \bar{m} = 0 ~~\text{for}~~ \lambda = r/s. 
\label{9}
\eeq
The bounds $\omega_\pm$ are easily found from the momentum and 
energy conservation, 
\beq
q = P_{\,\ket{R}}, 
\quad
\omega = E_{\ket{R}}.
\label{10}
\eeq
Indeed, it follows from Eqs. \eq{7}-\eq{10} that 
\beqa
q \!&=&\! \sum_{i=1}^s m(v_i-u) 
+ \,\sum_{j=1}^r \bar{m}(u-\bar v_j),
\label{11}\\
2(\omega - uq) 
\!&=& \! \sum_{i=1}^s m(v_i-u)^2 
- \,\sum_{j=1}^r \bar{m}(u-\bar v_j)^2.\qquad
\label{12}
\eeqa
Since $v_i-u\geq 0$ and $u-\bar v_j\geq 0$, \Eq{11} implies that for 
a given $q<2\pi \rho_0$ the velocities vary in the range
\beq
\begin{array}{cl}
\,u<v_i<v_0,
&
\quad v_0= u+q/m, 
\\
\bar v_0\!<\bar v_j\! < u,
&
\quad \bar v_0 = u-q/\bar{m}.
\end{array}
\label{13}
\eeq
The upper bound $\omega=\omega_+$ is reached when the velocity of 
one of the quasiparticles approaches $v_0$ while all the 
remaining quasiparticles/holes have velocities close to the plasma 
velocity $u$. \Eq{12} then gives  
\beq
\omega_+ -uq = \frac{m}{2}\,(v_0-u)^2 
= \frac{\,q^2}{2m}. 
\label{14}
\eeq
Similarly, the lower bound $\omega_-$ corresponds to the 
intermediate state $\ket{R}$ in which one of the quasiholes has 
velocity close to $\bar v_0$, while the velocities of all the remaining 
quasiparticles/holes are close to $u$, so that
\beq
\omega_- -uq = -\frac{\bar{m}}{2}\,(u-\bar v_0)^2 
= -\frac{\,q^2}{2\bar m} = - \lambda\frac{\,q^2}{2m}\,.
\label{15}
\eeq
The width of the region where $S\neq 0$ is then given by~\cite{drag}
\beq
\delta\omega =\omega_+-\omega_- 
= (\lambda+1) \frac{\,q^2}{2m}.
\label{16}
\eeq
\Eq{15} is valid as long as $\omega_-\geq 0$, i.e. for $q\leq 2\pi \rho_0$. 
At larger $q$ the function $\omega_-(q)$ is given by \Eq{15} 
with the replacement $q\to q-2\pi \rho_0 I$, where $I$ is the integer 
part of $q/2\pi \rho_0$. The corresponding intermediate state 
$\ket{R}$ has $I$ quasiholes with velocities approaching $-u$, 
one quasihole with velocity near $\bar v_0$ (given by \Eq{13} 
with the replacement $q\to q-2\pi \rho_0 I$), and all the remaining 
quasiparticles/holes moving with the plasma velocity $u$.

We turn now to the evaluation of the structure factor.
In the thermodynamic limit ($N\to\infty$, $\rho_0=\text{const}$) 
\Eq{1} can be rewritten as
\beq
S(q,\omega) =
q^2\!\!\!\int\! \prod_{i,j} dv_i d\bar v_j\,F_{s,r}\,
\delta\bigl(q-P_{\ket{R}}\bigr)\,
\delta\bigl(\omega-E_{\ket{R}}\bigr).
\label{17}
\eeq
The form-factor $F_{s,r}$ here 
is given  by 
\beq
F_{s,r} \propto 
\frac{
\prod_{i<i'}|v_{i}-v_{i'}|^{2\lambda}\!
\prod_{j<j'}|\bar v_{j}-\bar v_{j'}|^{2/\lambda}
}
{
\prod_{i,j}(v_i-\bar v_j)^2 
\prod_{i}\,\bigl(v_{i}^2 - u^2\bigl)^{1-\lambda}
\prod_{j}\,\bigl(u^2-\bar v_{j}^{2}\bigl)^{1-1/\lambda}
} \,.
\label{18}
\eeq
(note that $\prod dv_i d\bar v_j F_{s,r}$ is dimensionless).
This expression was conjectured in \cite{Haldane_conjecture} 
based on the results of Ref.~\cite{Altshuler} for $\lambda=1/2,2$~\cite{RMT}; 
the conjecture was proved in \cite{Ha} using properties of Jack polynomials. 
In writing \eq{18}, we omitted $\lambda$-dependent numerical coefficient~\cite{Ha}. 

For simplicity, we concentrate on the most interesting limit of small 
$q\ll \pi\rho_0$. In this limit $v_0-u,u-\bar v_0 \ll u$, 
see \Eq{13}, and one can approximate $v_{i}^2 - u^2\approx 2u(v_i - u)$,
$u^2-\bar v_{j}^{2}\approx 2u(u-\bar v_j)$ in \Eq{18}.
In view of the restriction \eq{13} on the velocities, it is 
convenient to switch in \Eq{17} to the new integration variables
\beq
x_i = \frac{v_i - u}{v_0-u}\,,
\quad
\bar x_j = \frac{u-\bar v_j}{u-\bar v_0}\,,
\label{19}
\eeq
which vary between $0$ and $1$. In terms of these variables
\beqa
\frac{S(q,\omega)}{S_0(q)} 
\!&=&\!\!
\int_0^1 \prod dx_i d\bar x_j\,F_{s,r}
\,\delta\Bigl(\tsum x_i + \tsum\bar x_j-1\Bigr)\qquad
\label{20}\\
&&\qquad
\times\,\delta\biggl(\!\tsum x_i^2 - \lambda\!\tsum\bar x_j^2
- (\lambda+1)\frac{\omega-uq}{\delta\omega}\biggr),
\nn
\eeqa
where $S_0(q)= m/q$ is the structure factor for 
noninteracting fermions $(\lambda=1)$, and the form-factor is
\beq
F_{s,r} \propto 
\frac{
\prod_{i<i'}|x_{i}-x_{i'}|^{2\lambda}\!
\prod_{j<j'}|\bar x_{j}-\bar x_{j'}|^{2/\lambda}
}
{
\prod_{i,j}(x_i + \lambda \bar x_j)^2 
\prod_{i}x_i^{1-\lambda}
\prod_{j}\bar x_j^{1-1/\lambda}
} \,.
\label{21}
\eeq

Evaluation of \Eq{20} simplifies considerably when 
$\omega\to\omega_\pm$. Consider, for example, the limit 
$\omega-\omega_-\ll\delta\omega$. In this limit the non-vanishing 
contributions to the integral in \eq{20} come from $r$ sectors  
of the $(r+s)$-dimensional integration space in which one of $\bar x_j$, 
say $\bar x_k$, is close to $1$, while all the other variables approach $0$.
To lowest order in $x_i,\bar x_{j\neq k},1-\bar x_k\ll 1$,
the second $\delta$-function in \Eq{20} can be replaced by 
\[
\delta(\ldots)\approx \,\delta\Bigl(2\lambda(1-\bar x_k) - (\lambda+1)
\frac{\,\omega-\omega_-}{\delta\omega\,\,}\Bigr),
\]
and the form-factor $F_{s,r}$ becomes
\[
F_{s,r}\Bigl(\{x_i\},\!\{\bar x_j\}\Bigr)
\approx
\lambda^{-2s}F_{s,r-1}\Bigl(\{x_i\},\!\{\bar x_j\}_{\!j\neq k}\Bigr)
\]
($F_{s,r-1}$ is given by \eq{21} with $j=k$ excluded). With these 
approximations, the integrations in \eq{20} are easily carried out and 
yield
\beq
\frac{S(q,\omega)}{S_0(q)} \propto 
\left[\frac{\,\omega-\omega_-}{\delta\omega~}\right]^{1/\lambda -1},
\quad
0<\omega-\omega_-\ll\delta\omega.
\label{22}
\eeq
Similarly, at $\omega_+ -\omega\ll\delta\omega$, the non-vanishing 
contributions to the integral in \Eq{20} come from $s$ sectors where
$1-x_k, x_{i\neq k}, \bar x_j\ll 1$, and one finds
\beq
\frac{S(q,\omega)}{S_0(q)} \propto 
\left[\frac{\omega_+\!-\omega}{\delta\omega}\right]^{\lambda-1},
\quad
0<\omega_+-\omega\ll\delta\omega.
\label{23}
\eeq

It should be emphasized that Eqs. \eq{22} and \eq{23} with $q$-independent 
exponents are valid for all $q\neq 2\pi\rho_0 I$, including $q\gtrsim\pi\rho_0$. 
In addition to $\omega_\pm(q)$, a smooth dependence on $q$ enters Eqs. 
\eq{22} and \eq{23} via the (omitted) prefactors; these $\omega$-independent 
prefactors tend to $1$ in the limit $\lambda\to 1$. According to Eqs. \eq{22} 
and \eq{23}, the structure factor diverges at $\omega\to\omega_-\,(\omega_+)$ 
for a repulsive (attractive) interaction, see Fig.~\ref{Fig1}(b); the divergencies 
are integrable for all physically meaningful (i.e. positive) values of 
$\lambda$~\cite{square-root}. 
Note that, formally, \Eq{23} can be obtained by replacing 
$|\omega-\omega_-|\to|\omega-\omega_+|$ and $\lambda\to1/\lambda$ 
in \Eq{22}. This correspondence is a manifestation of the 
particle$\,\leftrightarrow\,$hole, $\lambda\!\leftrightarrow \!1/\lambda$ 
duality of CSM~\cite{Polychronakos}. 

The power-law singularities in Eqs. \eq{22} and \eq{23} are similar to 
the familiar edge singularities in the x-ray absorption spectra in 
metals~\cite{Mahan}. Consider, for example, the limit $\omega\to\omega_+$. 
In this case the action of the operator $\rho_q^\dagger$ on the ground 
state creates a quasiparticle at $k\approx q$ with energy $\approx \omega_+$. 
This high-energy quasiparticle interacts with the Fermi sea, resulting in 
the excitation of multiple low-energy particle-hole pairs, just like the 
core hole does in the conventional x-ray edge singularity. 
The proliferation of the low-energy particle-hole pairs leads to the 
singularity in the response function~\cite{1DEG,Furusaki}. 

This analogy suggests that, just like in the case of the conventional 
edge singularity, the functional form of the dependence of the structure 
factor on $\omega-\omega_+$ can be captured by replacing the original 
model with the properly chosen effective Hamiltonian. For 
$\omega\to\omega_+$ and $q\ll \pi\rho_0$ it is sufficient to include in 
the effective Hamiltonian only the \textit{right-moving} single-particle states 
within two very narrow stripes (subbands) of momenta near $k=q$ and 
$k=0$~\cite{1DEG} to allow for both the creation of the high-energy 
quasiparticle and the low-energy particle-hole pairs. (Recall that for 
$q\!<\!\pi\rho_0$ the velocities of all quasiparticles/holes in the state 
$\rho_q^\dagger\ket{0}$ are \textit{positive}, see \Eq{13}; 
this would not be the 
case for a generic interaction~\cite{1DEG}). 

Upon introducing 
\[
\psi_r (x) = \!\!\sum_{|k|<k_0}\!\frac{e^{ikx}}{\sqrt{L}}\,\psi_k,
\quad
\psi_d(x) = \!\!\!\sum_{|k-q|<k_0}\!\!\!\!\frac{e^{i(k-q)x}}{\sqrt{L}}\,\psi_k,
\]
where $\psi_k$ annihilates a right-moving particle with momentum $k$ ($k=0$ 
at the Fermi level) and $k_0\ll q$ is a cutoff, the effective 
Hamiltonian can be written in the coordinate representation,
\beqa
H_+ \!&\!=\!&\!
\int\!dx\,\psi^\dagger_r\bigl(-iu\partial_x\bigr)\psi^\pdag_r
+ \!\int\!dx\, \psi_d^\dagger\bigl(\omega_+\! - iv_0\partial_x\bigr)\psi_d^\pdag
\nn\\
&\,&\qquad+\,U\!\!\int\!dx\,\rho_d(x)\rho_r(x).
\label{24}
\eeqa 
Here $\rho_{r,d} = \,\colon\!\psi^\dagger_{r,d}\psi^\pdag_{r,d}\colon\!\!$, 
where the colons denote the normal ordering. In \Eq{24} the nonlinearity of 
spectrum \eq{2} is encoded in the mismatch of the velocities
$v_0-u = q/m$, see \Eq{13}. The inter-subband interaction 
constant $U$ is set by the requirement that the two-particle scattering phase 
shift $\Theta = U/(v_0-u)$ for \Eq{24} is equal to that for CSM, 
$\Theta=(\lambda-1)\pi$~\cite{Sutherland}. This gives
\beq
U=(v_0-u)\Theta = (\lambda-1)\pi q/m,
\label{25}
\eeq
which for $|\lambda-1|\ll 1$ coincides with $V_0-V_q$~\cite{1DEG,Fourier}.

In terms of \Eq{24}, the structure factor is given by
\beq
S(q,\omega)\! = \!\int\!dx\,dt\,e^{i\omega t}
\bigl\langle b(x,t)b^\dagger(0,0)\bigr\rangle,
\quad
b^\dagger= \psi^\dagger_d\psi^\pdag_r.
\label{26}
\eeq
Note that the total number of $d$-particles $N_d = \int\!dx\,\rho_d(x)$ 
commutes with $H_+$ and that the entire $d$-subband lies above 
the Fermi level.  Hence, as far as the evaluation of \Eq{26} is concerned, 
$H_+$ can be further simplified by replacing 
$\psi_d(x)\to{\mathcal P}\psi_d(x){\mathcal P}$, 
where ${\mathcal P}$ is a projector onto states with $N_d=0,1$.
Obviously, the projected operators satisfy
$\rho_d(x)\psi_d(y) = 0$, and $\psi_d(x)\rho_d(y)=\delta(x-y)\psi_d(x)$,
which implies that $\bigl[\rho_d(x),\rho_d(y)\bigr] =0$.

We now bosonize the $\psi_r$-field according to~\cite{1D_books}
\[
\psi_r(x) = \sqrt{k_0}\,e^{i\varphi(x)},
~~
\bigl[\varphi(x),\varphi(y)\bigr]
= i\pi\sign(x-y),
\]
and apply a unitary transformation~\cite{1DEG} with generator 
$W = (\Theta/2\pi)\!\!\int\!dx\,\rho_d(x)\,\partial_x\varphi$.
The transformed Hamiltonian reads
\beq
\widetilde H_+ = e^{iW}H_+ e^{-iW} 
= H_0 + \delta H,
\label{27} 
\eeq
where
\beq
H_0 = \frac{u}{4\pi}\!\int\!dx\,(\partial_x\varphi)^2
+ \int\!dx\, \psi_d^\dagger\bigl(\omega_+\! - iv_0\partial_x\bigr)\psi_d^\pdag
\label{28} 
\eeq
and 
$\delta H = (v_0-u)(\Theta^2/4\pi)\!\int\!dx\,\rho_d^2(x)$.
It is easy to see that a state with a single $d$-particle is an eigenstate 
of $\delta H$, 
$\delta H\psi^\dagger_d(x)\!\ket{0}  
\propto 
(k_0/q)\delta\omega (\lambda-1)^2 \psi^\dagger_d(x)\!\ket{0}$.
Thus, when acting in the subspace with $N_d=0,1$, the second term in 
\Eq{27} results merely in a correction to $\omega_+$ in $H_0$, which 
for $k_0/q\ll 1$ can be safely neglected, i.e. $\widetilde H_+ \approx H_0$. 

The same unitary transformation applied to the operator $b^\dagger$ in 
\Eq{26} yields
\[
\widetilde b^\dagger(x)
= e^{iW}b^\dagger(x)\, e^{-iW}
= \sqrt{k_0}\,\psi^\dagger_d(x)\,e^{i\bigl(\!1+\Theta/2\pi\bigr)\varphi(x)}.
\]

Evaluation of the correlation function \eq{26} with quadratic Hamiltonian 
$H_0$ is now straightforward. The structure factor vanishes identically at
$\omega>\omega_+$, while at $\omega<\omega_+$ it is given by \Eq{23}. 
Thus, the outlined simplified description indeed reproduces the exact result 
\eq{23} with logarithmic accuracy. (The cutoff $k_0$ would enter \Eq{23} 
via a factor $q/k_0$ in the square brackets.) Note that the exponent in \Eq{23} 
is independent of $q$. This independence is a direct consequence of the fact 
that the phase shift $\Theta=\text{const}$ for inverse-square interaction.

Similar reasoning can be applied to the calculation of $S(q,\omega)$ 
at $\omega\to\omega_-$. In this case the $d$-subband lies near 
$k=-q$ well below the Fermi level and carries at most a single hole~\cite{1DEG}. 
After the particle-hole transformation $\psi^\pdag_{r,d}\to\psi^\dagger_{r,d}$
the corresponding effective Hamiltonian $H_-$ takes the form of \Eq{24} 
with replacements $\omega_+\to\omega_-$ and $v_0\to\bar v_0$. 
Evaluation of $S(q,\omega)$ [which is again given by \Eq{26}] proceeds 
similar to above and yields \Eq{22}. 

To conclude, in this Letter we evaluated the dynamic structure factor 
$S(q,\omega)$ of the Calogero-Sutherland model. Besides being of a 
fundamental interest for it's own sake, the detailed knowledge of the structure 
factor for interacting fermions with nonlinear dispersion is important for the 
description of a variety of effects associated with the particle-hole asymmetry. 

We found that $S(q,\omega)$ differs from zero in a finite interval of 
frequencies. At the borders of this interval $S(q,\omega)$ exhibits 
power-law singularities, analogous to the edge singularities in the x-ray 
absorption spectra of metals. Exploiting this analogy, we showed that 
the exact results \eq{22} and \eq{23} can be reproduced with logarithmic 
accuracy by replacing the original model with simple effective Hamiltonians. 
Remarkably, the analogy with the x-ray singularity, previously established 
for weak interaction only~\cite{1DEG}, is useful even when 
interactions are strong. Moreover, similar ideas can be applied 
to the evaluation of single-particle correlation functions~\cite{MP}. 

\begin{acknowledgments}
The author thanks Abdus Salam ICTP and William I. Fine 
Theoretical Physics Institute at the University of Minnesota 
for their hospitality and A. Abanov, B. Altshuler, F. Essler, 
L. Glazman, A. Kamenev, and P. Wiegmann for 
valuable discussions.
\end{acknowledgments}


\begin{thebibliography}{99}
\bibitem{structure_factor}
By the fluctuation-dissipation theorem,
$S(q,\omega) = 2\im\chi(q,\omega)$ at $\omega>0$ and $T=0$,
where $\chi$ is the corresponding retarded correlation function, 
see, e.g.,~\cite{Nozieres}.
\bibitem{Fourier}
The Fourier transform of \Eq{5} is easily found with the help of 
the identity $(\sin\alpha)^{-2} = - 2\sum_{n} |n|e^{2in\alpha}$ 
and reads $V_q = -\lambda(\lambda-1)\pi|q|/m$.
\end{thebibliography}
\end{document}